\documentclass[amsmath,amssymb,11pt]{article}
\usepackage{amsmath,amsfonts,latexsym,graphicx,amssymb}
\usepackage{amsfonts}
\usepackage{bm}
\date{}

\newcommand{\ot}{{\,\otimes\,}}
\newcommand{{\Cd}}{{\mathbb{C}^d}}

\def\<{\langle}
\def\>{\rangle}

\newtheorem{Proposition}{Proposition}

\newtheorem{remark}{Remark}
\numberwithin{equation}{section}

\begin{document}
\title{\bf A class of commutative dynamics of open quantum systems} \author{Dariusz
Chru\'sci\'nski  \\
Institute of Physics, Nicolaus Copernicus University,\\
Grudzi\c{a}dzka 5/7, 87--100 Toru\'n, Poland \\ \\
Andrzej Kossakowski \\
Dipartimento di Scienze Fisiche and MECENAS, \\ Universit\`a di
Napoli ``Federico II", I-80126 Napoli, Italy
\\ \\
Paolo Aniello, Giuseppe Marmo and Franco Ventriglia\\
Dipartamento di Scienze Fisiche, Universit\`a di Napoli ``Federico II'' \\
and Instituto Nazionale di Fisica Nucleare, Sezione di Napoli, \\
Complesso Universitario di Monte Sant Angelo,\\
Via Cintia, I-80126 Napoli, Italy }

\maketitle

\begin{abstract}
We analyze a class of dynamics of open quantum systems which is
governed by the dynamical map mutually commuting at different times.
Such evolution may be effectively described via spectral analysis of
the corresponding time dependent generators. We consider both
Markovian and non-Markovian cases.
\end{abstract}

\maketitle

\section{Introduction}
\setcounter{equation}{0}

The  dynamics of open quantum systems attracts nowadays increasing
attention \cite{OPEN}. It is very much connected to the growing
interest in controlling quantum systems and applications in modern
quantum technologies such as quantum communication, cryptography and
computation \cite{QIT}. The most popular approach is to use a
Markovian approximation and to consider a master equation
\begin{equation}\label{ME}
    \dot{A}_{t,t_0} = \mathcal{L}_t A_{t,t_0} \ , \ \ \ \ \ A_{t_0,t_0} = {\rm id}\ ,
\end{equation}
with time dependent generator $\mathcal{L}_t$. The above equation
gives rise to a quantum dynamical map (completely positive and trace
preserving) $A_t$ which in turn produces the evolution of a quantum
state $\rho_t = A_t \rho$. The corresponding generator
$\mathcal{L}_t$ has to satisfy well known condition
\cite{Gorini,Lindblad} (see also \cite{Alicki} for the detail
presentation) and the solution is given by the following formula
\begin{equation}\label{A-gen}
    A_{t,t_0} = \mathbb{T}\, \exp \left( \int_{t_0}^t \mathcal{L}_u \, du \right) \ ,
\end{equation}
where $\mathbb{T}$ denotes a chronological product. We stress that
the above formula has only a formal character since the evaluation
of its r.h.s. is in general not feasible. If the generator does not
depend on time $\mathcal{L}_t = \mathcal{L}$ then it simplifies to
\begin{equation}\label{A-com}
    A_{t,t_0} =  \exp \left(\mathcal{L}\, (t-t_0) \right) \ .
\end{equation}
Let us note that characteristic feature of (\ref{A-com}) is that
Markovian semigroup $A_t:= A_{t+t_0,t_0}$ is commutative, that is
\begin{equation}\label{com}
    A_t A_s = A_s A_t \ , \ \ \ \ s,t \geq 0 \ .
\end{equation}
It is no longer true for the general time dependent case
(\ref{A-gen}). The general Markovian evolution does satisfy the
inhomogeneous composition law
\begin{equation}\label{CL-0}
    A_{t,u} A_{u,s} = A_{t,s}\ ,
\end{equation}
for $t\geq u \geq s\geq t_0$, however, it is in general
noncommutative.

Non-Markovian evolution is much more difficult to analyze (see
\cite{Wilkie}--\cite{PRA-Saverio} for the recent papers). The local
master equation is replaced by the following equation
\begin{equation}\label{non-local}
\dot{A}_{t,t_0} = \int_{t_0}^t \mathcal{K}_{t-u}A_{u,t_0}\, du\ , \
\ \ \rho(t_0)=\rho_0\ ,
\end{equation}
in which quantum memory effects are taken into account through the
introduction of the memory kernel $\mathcal{K}_t$: this simply means
that the rate of change of the state $\rho(t)$ at time $t$ depends
on its history (starting at $t=t_0$). Recently, we proposed a
different approach \cite{PRL} which replaces the non-local equation
(\ref{non-local}) by the following local in time master equation
\begin{equation}\label{ME-non}
    \dot{A}_{t,t_0} = \mathcal{L}_{t-t_0} A_{t,t_0} \ , \ \ \ \ \ A_{t_0,t_0} = {\rm id}\
    .
\end{equation}
The price one pays for the local approach is that the corresponding
generator keeps the memory about the starting point `$t_0$'. This is
the very essence of non-Markovianity. Interestingly, this generator
might be highly singular, nevertheless,  the corresponding dynamics
is perfectly regular. Remarkably, singularities of generator may
lead to interesting physical phenomena like revival of coherence or
sudden death and revival of entanglement \cite{PRL}. Now, the formal
solution to (\ref{ME-non}) reads as follows
\begin{equation}\label{A-gen-non}
    A_{t,t_0} = \mathbb{T}\, \exp \left( \int_{0}^{t-t_0} \mathcal{L}_u \, du \right) \
    .
\end{equation}
It resembles very much Markovian dynamical map (\ref{A-gen}) and
again its r.h.s. has only formal character due to the presence of
the chronological operator. Note, however, important difference
between (\ref{A-gen}) and (\ref{A-gen}): the former does satisfy
composition law. The latter is homogeneous in time (depends upon the
difference $t-t_0$) but does not satisfy (\ref{CL-0}).

In the present paper we analyze a special case of commutative
dynamics, i.e. we generalize (\ref{com}) for time dependent
Markovian and non-Markovian dynamics. In this case formulae
(\ref{A-gen}) and (\ref{A-gen}) considerably simplify -- the
chronological product drops out and may compute the formula for the
dynamical map via spectral analysis.

\section{Preliminaries}
\setcounter{equation}{0}

Consider $d$-dimensional complex Hilbert space $\mathbb{C}^d$ and
let $\{e_0,\ldots,e_{d-1}\}$ be a fixed orthonormal basis. For any
$x,y \in \mathbb{C}^d$ denote by $\<x,y\>$ the corresponding scalar
product of $x$ and $y$. Let $M_d =
\mathcal{L}(\mathbb{C}^d,\mathbb{C}^d)$ denote a space of linear
operators in $\mathbb{C}^d$. Now, $M_d$ is equipped with the
Hilbert-Schmidt scalar product
\begin{equation}\label{}
    (a,b) := \sum_{k=0}^{d-1} \< ae_k,be_k\> = {\rm tr} (a^*b)\ ,
\end{equation}
where $a^* : \mathbb{C}^d \rightarrow \mathbb{C}^d$ is defined by
\begin{equation}\label{}
    \<a^* x,y\> = \<x,ay\> \ ,
\end{equation}
for arbitrary $x,y \in \mathbb{C}^d$. Finally, let us introduce the
space $\mathcal{L}(M_d,M_d)$ of linear maps $A : M_d \rightarrow
M_d$. For any $A \in \mathcal{L}(M_d,M_d)$ one defines a dual map
$A^\# \in \mathcal{L}(M_d,M_d)$ by
\begin{equation}\label{}
    (A^\# a,b) = (a,Ab) \ ,
\end{equation}
for arbitrary $a,b \in M_d$. Note, that if the dual map $A^\#$ is
unital, i.e. $A^\# \mathbb{I}_d = \mathbb{I}_d$, then $A$ is trace
preserving. It is clear that $\mathcal{L}(M_d,M_d)$ defines
$d^2\times d^2$ complex Hilbert space equipped with the following
inner product
\begin{equation}\label{}
    \<\< A,B\>\> = \sum_{\alpha=0}^{d^2-1} (Af_\alpha,Bf_\alpha) =
    \sum_{\alpha=0}^{d^2-1} {\rm tr}\, [(Af_\alpha)^*
    (Bf_\alpha)] \ ,
\end{equation}
for any $A,B \in \mathcal{L}(M_d,M_d)$. In the above formula
$f_\alpha$ denote an orthonormal basis in $M_d$. Let us observe that
in $\mathcal{L}(M_d,M_d)$ one constructs two natural orthonornal
basis
\begin{equation}\label{}
    F_{\alpha\beta} \, :\, M_d \longrightarrow M_d\ ,
\end{equation}
and
\begin{equation}\label{}
    E_{\alpha\beta} \,:\, M_d \longrightarrow M_d\ ,
\end{equation}
defined as follows
\begin{equation}\label{f-ab}
    F_{\alpha\beta} a = f_\alpha a f_\beta^*\ ,
\end{equation}
and
\begin{equation}\label{e-ab}
    E_{\alpha\beta} a = f_\alpha (f_\beta,a)\ ,
\end{equation}
for any $a \in M_d$. One easily proves
\begin{equation}\label{}
    \<\< F_{\alpha\beta},F_{\mu\nu} \>\> =   \<\< E_{\alpha\beta},E_{\mu\nu}
    \>\> = \delta_{\alpha\mu}\delta_{\beta\nu} \ .
\end{equation}
Moreover, the following relations are satisfied
\begin{equation}\label{}
 \sum_{\alpha=0}^{d^2-1} F_{\alpha\alpha}\, a = \mathbb{I}_d \, {\rm tr} a\ ,
\end{equation}
and
\begin{equation}\label{}
 \sum_{\alpha=0}^{d^2-1} E_{\alpha\alpha}\, a = a \ .
\end{equation}
\begin{remark}{\em Note, that representing a linear map $A$ in the
basis $F_{\alpha\beta}$
\begin{equation}\label{}
    A = \sum_{\alpha,\beta} a_{\alpha\beta}\, F_{\alpha\beta} \ ,
\end{equation}
with
\begin{equation}\label{}
    a_{\alpha\beta} = \<\< A,F_{\alpha\beta}\>\>\ ,
\end{equation}
one has a simple criterion for complete positivity of $A$: a map $A$
is complete positive if and only if the corresponding $d^2 \times
d^2$ matrix $||a_{\alpha\beta}||$ is semipositive definite. On the
other hand the $E$-representation
\begin{equation}\label{}
    A = \sum_{\alpha,\beta} a'_{\alpha\beta}\, E_{\alpha\beta} \ ,
\end{equation}
with
\begin{equation}\label{}
    a'_{\alpha\beta} = \<\< A,E_{\alpha\beta}\>\>\ ,
\end{equation}
does not give any simple criterion for complete positivity. Note,
however, that $E$-representation is well suited for the composition
of maps. If
\begin{equation}\label{}
    B = \sum_{\alpha,\beta} b'_{\alpha\beta}\, E_{\alpha\beta} \ ,
\end{equation}
with
\begin{equation}\label{}
    b'_{\alpha\beta} = \<\< B,E_{\alpha\beta}\>\>\ ,
\end{equation}
then the map $C = A \circ B$ gives rise to the following
representation
\begin{equation}\label{}
    C = \sum_{\alpha,\beta} c'_{\alpha\beta}\, E_{\alpha\beta} \ ,
\end{equation}
where the matrix $c' = a' \cdot b'$.
 }
\end{remark}

 Consider now a
linear map $A$ from $\mathcal{L}(M_d,M_d)$ and let us assume that
$A$ is diagonalizable, that is, it gives rise to the Jordan
representation with 1-dimensional Jordan blocks. One has
\begin{equation}\label{}
    A = V \, D\, V^{-1}\ ,
\end{equation}
where $D$ is diagonal. It means that there exists an orthonormal
basis $f_\alpha$ in $M_d$  such that
\begin{equation}\label{}
    \<\< f_\alpha,D f_\beta\>\> = d_\alpha \delta_{\alpha\beta}\ ,
\end{equation}
with $d_\alpha \in \mathbb{C}$. It shows that
\begin{equation}\label{}
    D = \sum_{\alpha = 0}^{d^2-1} d_\alpha\, P_\alpha \ ,
\end{equation}
where
\begin{equation}\label{}
    P_\alpha a = f_\alpha (f_\alpha,a)\ , \ \ \  a \in M_d\ .
\end{equation}
Note, that a set $P_\alpha$ defines a family of orthogonal
projectors
\begin{equation}\label{}
  P_\alpha P_\beta =
\delta_{\alpha\beta} P_\alpha\ ,
\end{equation}
together with
\begin{equation}\label{}
\sum_{\alpha = 0}^{d^2-1}  P_\alpha = {\rm id} \ ,
\end{equation}
where ${\rm id}$ denotes an identity map in $\mathcal{L}(M_d,M_d)$.
Hence, one obtains the following representation of $A$
\begin{eqnarray}\label{}
    A a &=& V \, D\, V^{-1} a = \sum_{\alpha = 0}^{d^2-1} d_\alpha\,
    VP_\alpha V^{-1} a \nonumber \\ &=& \sum_{\alpha = 0}^{d^2-1} d_\alpha\,
    V f_\alpha (f_\alpha, V^{-1} a) = \sum_{\alpha = 0}^{d^2-1} d_\alpha\,
    V f_\alpha (V^{-1\#}f_\alpha,  a) \ .
\end{eqnarray}
Let us define new basis
\begin{equation}\label{}
    g_\alpha := Vf_\alpha\ , \ \ \ \ h_\alpha := V^{-1\#} f_\alpha\ .
\end{equation}
Note $g_\alpha$ and $h_\alpha$ define a pair of bi-orthogonal (or
damping \cite{damping}) basis in $M_d$
\begin{equation}\label{}
    (g_\alpha,h_\beta) = (Vf_\alpha,V^{-1\#}f_\beta) =
    (f_\alpha,f_\beta) = \delta_{\alpha\beta}\ .
\end{equation}
Finally, one obtains the following spectral representation of the
linear map $A$
\begin{equation}\label{}
     A = \sum_{\alpha = 0}^{d^2-1} d_\alpha\, \widetilde{P}_\alpha\
     ,
\end{equation}
where
\begin{equation}\label{}
 \widetilde{P}_\alpha a := g_\alpha (h_\alpha,a) \  , \ \ \ \  a \in
 M_d \ .
\end{equation}
Note, that a set $\widetilde{P}_\alpha$ satsfies
\begin{equation}\label{}
  \widetilde{P}_\alpha \widetilde{P}_\beta =
\delta_{\alpha\beta} \widetilde{P}_\alpha\ ,
\end{equation}
together with
\begin{equation}\label{}
\sum_{\alpha = 0}^{d^2-1}  \widetilde{P}_\alpha = {\rm id} \ .
\end{equation}
However, contrary to $P_\alpha$ operators $\widetilde{P}_\alpha$ are
not Hermitian, i.e. $\widetilde{P}_\alpha^\# \neq
\widetilde{P}_\alpha$
\begin{equation}\label{}
\widetilde{P}_\alpha^\# a := h_\alpha (g_\alpha,a) \  , \ \ \ \  a
\in  M_d \ .
\end{equation}
It shows that $\widetilde{P}_\alpha$ are not projectors unless
$g_\alpha = h_\alpha$. The corresponding spectral representation of
the dual map $A^\#$ reads as follows
\begin{equation}\label{}
     A^\# = \sum_{\alpha = 0}^{d^2-1} \overline{d_\alpha}\, \widetilde{P}_\alpha^\#\
     ,
\end{equation}
where $\overline{x}$ stands for the complex conjugation of the
complex number $x$. Hence, one obtains the following family of
eigenvectors
\begin{equation}\label{}
    A g_\alpha = d_\alpha g_\alpha \ , \ \ \ \ A^\# h_\alpha =
    \overline{d_\alpha}     h_\alpha \ .
\end{equation}
Consider for example a special case with $V=U$ and $U$ is a unitary
operator in $M_d$. One has $U^\# = U^{-1}$ and hence $V^{-1\#} = U$.
One obtains
\begin{equation}\label{}
    g_\alpha \equiv h_\alpha = U f_\alpha\ ,
\end{equation}
and hence $\widetilde{P}_\alpha^\# = \widetilde{P}_\alpha$. Note
that
\begin{equation}\label{}
A = \sum_{\alpha = 0}^{d^2-1} {d_\alpha}\, \widetilde{P}_\alpha\ , \
\ \ \  A^\# = \sum_{\alpha = 0}^{d^2-1} \overline{d_\alpha}\,
\widetilde{P}_\alpha \ ,
\end{equation}
which implies that the super-operator $A$ is normal
\begin{equation}\label{}
   AA^\# = A^\#A \ .
\end{equation}

\section{How to generate commutative dynamics}
\setcounter{equation}{0}

Consider a family of Markovian semigroups $A^{(k)}_t$ defined by
\begin{equation}\label{}
A^{(k)}_t = e^{t L_k} \ , \ \ \ k=1,\ldots,n \ ,
\end{equation}
where $L_k$ are the corresponding generators. Suppose that $L_k$ are
mutually commuting and define
\begin{equation}\label{}
    A_{t,t_0} = \sum_{k=1}^n p_k(t-t_0) \, A_{t-t_0}^{(k)}\ ,
\end{equation}
where $p_k(t)$ denotes time dependent probability distribution:
$p_k(t) \geq 0$ and $p_1(t) + \ldots + p_n(t) =1$. Let us observe
that $A_{t,t_0}$ defines a commutative non-Markovian evolution
satisfying local in time Master Equation \cite{PRL}
\begin{equation}\label{}
  \dot{A}_{t,t_0} = \mathcal{L}_{t-t_0} A_{t,t_0}\ , \ \ \ \ \ A_{t_0,t_0} = {\rm id}\
  .
\end{equation}
To find the non-Markovian generator $\mathcal{L}_t$ let us assume
the following spectral representation of $L_k$
\begin{equation}\label{}
    L_k \rho = \sum_\alpha \lambda_\alpha^{(k)} g_\alpha {\rm
    tr}(h_\alpha^* \rho) \ .
\end{equation}
One obtains
\begin{equation}\label{}
    \mathcal{L}_t \rho = \sum_\alpha \mu_\alpha(t) g_\alpha {\rm
    tr}(h_\alpha^* \rho) \ ,
\end{equation}
with
\begin{equation}\label{}
    \mu_\alpha(t) = \frac{ \sum_k p_k(t) \lambda_\alpha^{(k)} e^{
    \lambda_\alpha^{(k)} t} }{ \sum_j p_j(t) e^{
    \lambda_\alpha^{(j)} t}} \ .
\end{equation}
Hence, the solution $A_t$ has the following form
\begin{equation}\label{}
    A_{t,t_0}\rho  = \sum_\alpha \exp\left( \int_0^{t-t_0} \mu_\alpha(u)\, du \right)
    \, g_\alpha {\rm tr}(h_\alpha^* \rho) \ .
\end{equation}
Actually, one can easily generate a family of commuting generators
$L_1,\ldots,L_n$.  Suppose one is given a Markovian generator $L$ of
a unital semigroup $A_t = e^{L t}$. Denote by $\widehat{A}_s$ the
Laplace transform of $A_t$
\begin{equation}\label{}
    \widehat{A}_s = \int_0^\infty e^{-st} A_t\, dt\, = \frac{1}{s- L}\ .
\end{equation}
It is evident that for $s > 0$, $\widehat{A}_s$ is completely
positive. Moreover,
\begin{equation}\label{}
    \Phi^{(0)}_s := s \widehat{A}_s\ ,
\end{equation}
is unital. Indeed, one has
\begin{equation}\label{}
\Phi^{(0)}_s\, \mathbb{I} = s\int_0^\infty e^{-st} dt\, \mathbb{I} =
\mathbb{I}\ .
\end{equation}
Now, let us define
\begin{equation}\label{}
    \Phi^{(k)}_s := \frac{s^{k+1}}{k!} (-1)^k\, \frac{d^k}{ds^k} \widehat{A}_s\
    .
\end{equation}
One gets
\begin{equation}\label{}
\Phi^{(k)}_s = \frac{s^{k+1}}{k!}\, \int_0^\infty e^{-st} t^k A_t\,
 dt\, = \frac{s^{k+1}}{(s- L)^{k+1}}\ .
\end{equation}
It is clear that for $s >0\,$, $\Phi^{(k)}_s$ is completely positive
and unital. Therefore, for any integer $k$ and $s > 0$ one obtains
the following Markovian generator
\begin{equation}\label{}
    L^{(k)}_s = \Phi^{(k)}_s - {\rm id} \ .
\end{equation}
Hence, fixing $s$, one arrives at $L_k := L^{(k)}_s$.

Let us observe that the construction of the commutative
$(s,k)$--family $L^{(s)}_k$ may be used to construct a huge family
of commuting time dependent generators. Note, that taking a discrete
family of function $f_k$
\[  f_k\ :\ \mathbb{R}_+ \times
\mathbb{R}_+\  \longrightarrow \ \mathbb{R}_+  \ , \] one may define
\begin{equation}\label{}
    \mathcal{L}_t[\mathbf{f}] := \sum_k \int_0^\infty f_k(t,s) L^{(k)}_s \, ds\
    ,
\end{equation}
where we used a compact notation $\mathbf{f} =(f_1,f_2,\ldots)$.  It
is clear from the construction that
$[\mathcal{L}_t[\mathbf{f}],\mathcal{L}_s[\mathbf{f}]]=0$, and hence
one easily find for the evolution
\begin{equation}\label{}
    A_{t,t_0}[\mathbf{f}] = \exp\left( \int_0^{t-t_0} {\cal L}_u[\mathbf{f}]\, du \right)\ .
\end{equation}
defines a family of commuting time dependent Markovian generators.

\section{A class of commutative dynamics of stochastic classical systems}
\setcounter{equation}{0}

\subsection{Markovian classical dynamics}

Consider the dynamics of a stochastic $d$-level system described by
a probability distribution $p=(p(0),\ldots,p(d-1))$. Its time
evolution is defined by
\begin{equation}\label{}
    p_t(m) = \sum_{n=0}^{d-1} T_{t,t_0}(m,n) p_0(n) \ ,
\end{equation}
where $T_{t,t_0}(n,m)$ is a stochastic matrix satisfying  the
following time-dependent master equation
\begin{equation}\label{C1}
    \dot{T}_{t,t_0} = {L}_t \,T_{t,t_0}\ , \ \ \ T_{t_0,t_0} = \mathbb{I}_d \
    ,
\end{equation}
that is
\begin{equation}\label{C1}
    \dot{T}_{t,t_0}(m,n) = \sum_{k=0}^{d-1}{L}_t(m,k) \,T_{t,t_0}(k,n)\ , \ \ \ \ \ \ T_{t_0,t_0}(m,n) = \delta(n,m)\
    .
\end{equation}
 Let us assume that $L_t$ defines a commuting family of $d \times d$ matrices, i.e.
\begin{equation}\label{}
    \sum_{k= 0}^{d-1}  \, L_t(m,k)L_u(k,n) = \sum_{k= 0}^{d-1}  \,
    L_u(m,k)L_t(k,n)\ ,
\end{equation}
for any $t,u \geq t_0 $. A particular example of commutative
dynamics is provided by circulant generators. Let us recall that a
$d \times d$  matrix $L(m,n)$ is circulant \cite{Gantmacher} if
\begin{equation}\label{}
    L(m,n) = a(m-n)\ , \ \ \ {\rm mod}\ d\ ,
\end{equation}
that is $L$ is defined in terms of a single vector
$a=(a(0),\ldots,a(d-1))$.

\begin{Proposition} Circulant matrices define a commutative
subalgebra of $M_d$. Hence, if $L$ and $L'$ are circulant then $L''
= LL' = L'L$ is circulant. Moreover, if
\[   L(i,j) = a(i-j)\ ,  \ \ \   L'(i,j) = a'(i-j)\ ,\ \ \ L''(i,j) = a''(i-j) \ , \ \ \ {\rm mod}\ d\ ,\]
then
\begin{equation}\label{}
    a'' = a \ast a' \ ,
\end{equation}
where $a \ast a'$ denotes a discrete convolution in $\mathbb{Z}_d$,
i.e.
\begin{equation}\label{}
    a''(n) = \sum_{k=0}^{d-1} a(n-k)\, a'(k) \ .
\end{equation}
\end{Proposition}
Therefore, multiplication of circulant matrices induces convolution
of defining $d$-vectors. Interestingly, spectral properties of
circulant matrices are governed by the following

\begin{Proposition} The eigenvalues $l_m$ and eigenvectors $\psi_m$ of a circulant
matrix
\begin{equation}\label{}
    L\, \psi_m = l_m\, \psi_m \ ,
\end{equation}
read as follows:
\begin{equation}\label{}
l_m = \sum_{k=0}^{d-1} a_k \lambda^{mk} \ ,
\end{equation}
and
\begin{equation}\label{}
    (\psi_m)_n = \frac{1}{\sqrt{d}}\, \lambda^{mn}\ ,
\end{equation}
where
\begin{equation}\label{lambda}
    \lambda = e^{2\pi i /d } \ .
\end{equation}
\end{Proposition}
Let us observe that the Kolmogorov conditions for the stochastic
circulant generator $L_t$
give rise to the following condition upon the time dependent vector
$a_t(m)$:
\begin{enumerate}
\item $\ a_t(m) \geq\ 0\,$, for $\ m\neq 0$
\item $\ a_t(0) < 0\, $,
\item $\ \sum_{m} a_t(m) = 0\, $,
\end{enumerate}
for $t \geq t_0$. Now, it is clear that the solution to (\ref{C1})
\begin{equation}\label{}
    T_{t,t_0} = \exp\left( \int_{t_0}^t L_u du \right) \ ,
\end{equation}
defines a circulant stochastic matrix. Hence
\begin{equation}\label{}
    T_{t,t_0}(m,n) =: P_{t,t_0}(m-n) \ ,
\end{equation}
defines a time-dependent stochastic vector $P_{t,t_0}(m)$. Note that
\begin{equation}\label{}
    p_t(m) = \sum_{n=0}^{d-1} T_{t,t_0}(m,n) p_0(n) = \sum_{n=0}^{d-1} P_{t,t_0}(m-n) p_0(n)\ ,
\end{equation}
and hence
\begin{equation}\label{}
    p_t = P_{t,t_0} \ast p_0 \ .
\end{equation}
One obtains from (\ref{C1})
\begin{equation}\label{C2}
    \frac{dP_{t,t_0}(m)}{dt} = \sum_{k=0}^{d-1} a_t(m-k)\,  P_{t,t_0}(k) \
    , \ \ \ \ P_{t_0,t_0}(m) = \delta_{m0} \ ,
\end{equation}
which can be rewritten  in terms of discrete convolution
\begin{equation}\label{C3}
    \dot{P}_{t,t_0} = a_t \ast P_{t,t_0}\ , \ \ \ \  \ P_{t_0,t_0}=e\
    ,
\end{equation}
where `$e$' corresponds to the distribution concentrated at $0$,
i.e. $e(m) = \delta_{m0}$.

\begin{Proposition}
A convex set $\mathcal{P}_d$ of probabilistic $d$-vectors  defines a
semigroup with respect to the discrete convolution. The unit element
$e=(1,0,\ldots,0)$ satisfies
\[  P \ast e = e \ast P = P \ ,  \]
for all $P \in \mathcal{P}_d$.
\end{Proposition}
To solve (\ref{C3}) one transform it via  discrete Fourier transform
to get
\begin{equation}\label{C4}
    \frac{d \widetilde{P}_{t,t_0}(m)}{dt} =  \widetilde{a}_t(m)\,
    \widetilde{P}_{t,t_0}(m)\ , \ \ \ \ \ \ \widetilde{P}_{t_0,t_0}(m)=1\
    ,
\end{equation}
where
\begin{equation}\label{}
    \widetilde{x}(n) = \sum_{k=0}^{d-1} \lambda^{nk}\,
    x(k) \ ,
\end{equation}
and the inverse transform reads
\begin{equation}\label{}
    x(k) = \frac{1}{d} \sum_{n=0}^{d-1} \lambda^{-nk}\,
    \widetilde{x}(n) \ .
\end{equation}
The solution of (\ref{C4}) reads as follows
\begin{equation}\label{C5}
    \widetilde{P}_{t,t_0}(m) = \exp\left( \int_{t_0}^t \widetilde{a}_u(m)du \right)\
    ,
\end{equation}
and hence one obtains for the stochastic vector $P_{t,t_0}(m)$
\begin{equation}\label{C5a}
{P}_{t,t_0}(m) = \frac 1d\, \sum_{k=0}^{d-1} \lambda^{-mk} \,
\exp\left( \int_{t_0}^t \widetilde{a}_u(m)du \right) \ .
\end{equation}
It is clear that $P_{t,t_0}$ satisfies the following composition law
\begin{equation}\label{CL-1}
    P_{t,s} \ast P_{s,u} = P_{t,u} \ ,
\end{equation}
or equivalently
\begin{equation}\label{}
    \widetilde{P}_{t,s} \cdot \widetilde{P}_{s,u} = \widetilde{P}_{t,u} \ ,
\end{equation}
 for all $t\geq s \geq u$. In particular when $a(n)$
does not depend on time then (\ref{C5}) simplifies to
\begin{equation}\label{C6}
    \widetilde{P}_{t,t_0}(m) = \exp\left(\widetilde{a}(m)[t-t_0]\right)\
    ,
\end{equation}
and hence  1-parameter semigroup $P_{t-t_0} := P_{t,t_0}$ satisfies
homogeneous composition law
\begin{equation}\label{CL-0}
    P_{t} \ast P_{s} = P_{t+s} \ ,
\end{equation}
or equivalently
\begin{equation}\label{}
    \widetilde{P}_{t} \cdot \widetilde{P}_{s} = \widetilde{P}_{t+s} \ ,
\end{equation}
for all $t\geq s \geq t_0$.

\subsection{Non--Markovian classical dynamics}

Consider now the non-Markovian case governed by the following local
in time master equation
\begin{equation}\label{C3-n}
    \dot{P}_{t,t_0} = a_{t-t_0} \ast P_{t,t_0}\ , \ \ \ \  \ P_{t_0,t_0}=e\
    .
\end{equation}
One easily obtain for the solution
\begin{equation}\label{C5a-n}
{P}_{t,t_0}(m) = \frac 1d\, \sum_{k=0}^{d-1} \lambda^{-mk} \,
\exp\left( \int_{0}^{t-t_0} \widetilde{a}_u(m)du \right) \ .
\end{equation}
Note the crucial difference between (\ref{C5a}) and (\ref{C5a-n}).
The former defines inhomogeneous semigroup whereas the latter is
homogeneous in time (depends upon the difference `$t-t_0$') but does
not define a semigroup, i.e. does not satisfy the composition law
(\ref{CL-1}).

Let us analyze conditions for $a_\tau$ which do guarantee that
${P}_{t,t_0}$ defined in (\ref{C5a-n}) is a probability vector, that
is,
\[ {P}_{t,t_0}(m) \geq 0\ , \ \ \ \ \ \sum_{m=0}^{d-1}
{P}_{t,t_0}(m) = 1 \ , \] for all $t \geq t_0$. It is clear from
(\ref{C3-n}) that $a(\tau)$ has to satisfy
\begin{equation}\label{KOL-1}
\int_{0}^{\tau} {a}_u(m)du \geq 0 \ ,
\end{equation}
for $m > 0$, and
\begin{equation}\label{KOL-2}
\sum_{m=0}^{d-1} \int_{0}^{\tau} {a}_u(m)du = 0 \ ,
\end{equation}
which implies that
\begin{equation}\label{KOL-3}
\int_{0}^{\tau} {a}_u(0)du < 0 \ ,
\end{equation}
for all $\tau \geq 0$. These conditions generalize Kolmogorov
conditions in the inhomogeneous Markovian case. We stress, that
$a_u(m)$ needs not be positive (for $m>0$). One has a weaker
condition (\ref{KOL-1}). Note, that if ${a}_t(m) \geq 0$ for $m>0$,
then $\int_{0}^{\tau} {a}_u(m)du$ defines a monotonic function of
time and hence the non-Markovian relaxation $\exp( \int_{0}^{\tau}
{a}_u(m)du)$ is monotonic in time as well.

Finally, let us consider the corresponding nonlocal equation
\begin{equation}\label{CM-n}
    \dot{P}_{t,t_0} = \int_{t_0}^t K_{t-u} \ast P_{u,t_0}\, du\ , \ \ \ \  \ P_{t_0,t_0}=e\
    ,
\end{equation}
with the memory kernel $K_{t-u}$. Note, that we already know
solution represented by (\ref{C5a-n}) but still do not know the
memory kernel $K$. Performing discrete Fourier transform one gets
from (\ref{CM-n})
\begin{equation}\label{CM-n}
    \dot{\widetilde{P}}_{t,t_0}(m) = \int_{t_0}^t \widetilde{K}_{t-u}(m) \widetilde{P}_{u,t_0}(m)\, du\ ,
    \ \ \ \  \ \widetilde{P}_{t_0,t_0}(m)=1 \ .
\end{equation}
Define the time-dependent vector
\begin{equation}\label{}
    f_t(m) = \widetilde{a}_t(m) \, \exp\left( \int_0^t
    \widetilde{a}_u(m)du \right) \ ,
\end{equation}
then following \cite{KR} one obtains
\begin{equation}\label{}
    \widehat{\widetilde{K}}_s(m) = \frac{s \widehat{f}_s(m)}{1 +
    \widehat{f}_s(m)} \ ,
\end{equation}
where $\widehat{x}_s$ denote the Laplace transform of $x_t$.
Clearly, the problem of performing the inverse Laplace transform
$\widehat{\widetilde{K}}_s(m) \longrightarrow {\widetilde{K}}_t(m)$
is in general not feasible. Hence, the memory kernel remains
unknown. Nevertheless, the solution is perfectly known.

\begin{remark}
{\em Note that a stochastic map $p_0 \rightarrow p_t= T_{t,t_0}p_0$
may be rewritten in a `quantum fashion' as follows. Any probability
distribution $p=(p(0),\ldots,p(d-1))$ gives rise to a diagonal
density matrix
\begin{equation}\label{}
    \rho = \sum_{n=0}^{d-1} p(n) e_{nn} \ ,
\end{equation}
and the map $\rho_0 \rightarrow \rho_t$ reads as follows
\begin{equation}\label{}
    \rho_t = \sum_{m,n=0}^{d-1} T_{t,t_0}(m,n)\, e_{mm}\, \rho_0 \, e_{nn}\
    .
\end{equation}
}
\end{remark}

\subsection{Dynamics of composite systems}

Consider now dynamics of $N$-partite system living in
$\mathbb{Z}^N_d = \mathbb{Z}_d \times \ldots \times \mathbb{Z}_d\,$.
Let $\mathbf{n}=(n_1,\ldots,n_N)$, with $n_k \in \mathbb{Z}_d\,$ and
let
\begin{equation}\label{}
    P_{t,t_0} \ :\ \mathbb{Z}^N_d \ \longrightarrow \ [0,1]\ ,
\end{equation}
be a probability vector living on $\mathbb{Z}^N_d\,$ satisfying the
following Markovian master equation
\begin{equation}\label{}
    \dot{P}_{t,t_0} = a_t \ast P_{t,t_0}\ , \ \ \ \  \ P_{t_0,t_0}=e\
    ,
\end{equation}
where `$e$' is defined by
\begin{equation}\label{}
    e(\mathbf{n}) = \delta_{\mathbf{n}\mathbf{0}} := \delta_{n_10}
    \ldots \delta_{n_N0}\ .
\end{equation}
Now, performing the discrete Fourier transform one gets
\begin{equation}\label{C4-multi}
    \frac{d \widetilde{P}_{t,t_0}(\mathbf{m})}{dt} =  \widetilde{a}_t(\mathbf{m})\,
    \widetilde{P}_{t,t_0}(\mathbf{m})\ , \ \ \ \ \ \ \widetilde{P}_{t_0,t_0}(\mathbf{m})=1\
    ,
\end{equation}
where
\begin{equation}\label{}
    \widetilde{x}(\mathbf{m}) = \sum_{\mathbf{k}} \lambda^{\mathbf{m}\mathbf{k}}\,
    x(\mathbf{k}) \ ,
\end{equation}
and the inverse transform reads
\begin{equation}\label{}
    x(\mathbf{k}) = \frac{1}{d^N} \sum_{\mathbf{m}} \lambda^{-\mathbf{m}\mathbf{k}}\,
    \widetilde{x}(\mathbf{n}) \ .
\end{equation}
The solution of (\ref{C4-multi}) reads as follows
\begin{equation}\label{}
    \widetilde{P}_{t,t_0}(\mathbf{m}) = \exp\left( \int_{t_0}^t \widetilde{a}_u(\mathbf{m})du \right)\
    ,
\end{equation}
and hence one obtains for the stochastic vector
$P_{t,t_0}(\mathbf{m})$
\begin{equation}\label{}
{P}_{t,t_0}(\mathbf{m}) = \frac{1}{d^N}\, \sum_{\mathbf{k}}
\lambda^{-\mathbf{m}\mathbf{k}} \, \exp\left( \int_{t_0}^t
\widetilde{a}_u(\mathbf{m})du \right) \ .
\end{equation}
It is clear that $P_{t,t_0}$ satisfies the inhomogeneous composition
law (\ref{CL-1}).  If $a_{\mathbf{m}}$ is time independent then
${P}_{t,t_0}$ defines 1-parameter semigroup $P_{\tau} :=
P_{\tau+t_0,t_0}$ satisfying homogeneous composition law
(\ref{CL-0}).

Note, that in the case of non-Markovian dynamics one has
\begin{equation}\label{}
    \dot{P}_{t,t_0} = a_{t-t_0} \ast P_{t,t_0}\ , \ \ \ \  \ P_{t_0,t_0}=e\
    ,
\end{equation}
giving rise to the following solution
\begin{equation}\label{}
{P}_{t,t_0}(\mathbf{m}) = \frac{1}{d^N}\, \sum_{\mathbf{k}}
\lambda^{-\mathbf{m}\mathbf{k}} \, \exp\left( \int_{0}^{t-t_0}
\widetilde{a}_{\mathbf{m}}(u)du \right) \ .
\end{equation}
The non-Markovian dynamics is time homogeneous but does not satisfy
(\ref{CL-1}).

\section{A class of commutative quantum dynamics}
\setcounter{equation}{0}

Consider now an abelian group $\mathbb{Z}_d \times \mathbb{Z}_d$.
Equivalently, one may consider a cyclic toroidal lattice
$\mathbb{T}_d \times \mathbb{T}_d$, where
\begin{equation}\label{}
    \mathbb{T}_d = \{ \lambda^m \ , \ m=0,1,\ldots,d-1\, \} \ ,
\end{equation}
which is an abelian multiplicative group. Let us define the
following representation of $\mathbb{T}_d \times \mathbb{T}_d$ in
$M_d$:
\begin{equation}\label{pro-REP}
    \mathbb{Z}_d \times \mathbb{Z}_d \ni (m,n) \ \longrightarrow\ u_{mn} \in M_d \ ,
\end{equation}
where $u_{mn}$ are unitary matrices defined as follows
\begin{equation}\label{}
    u_{mn} e_k = \lambda^{mk} e_{n+k} \ ,
\end{equation}
where $\{e_0,\ldots,e_{d-1}\}$ denotes an orthonormal basis in
$\mathbb{C}^d$, and $\lambda$ stands for $d$th root of identity (see
formula (\ref{lambda})).

\begin{Proposition}
Matrices $u_{mn}$ satisfy
\begin{eqnarray}
  u_{mn} u_{rs} &=& \lambda^{ms}\, u_{m+r,n+s} \ , \\
  u^*_{mn} &=& \lambda^{mn}\, u_{-m,-n} \ ,
\end{eqnarray}
and the following orthogonality relations
\begin{equation}\label{}
    {\rm tr}( u_{mn}^* u_{kl}) = d\, \delta_{mk} \delta_{nl}\ .
\end{equation}

\end{Proposition}
Hence, formula (\ref{pro-REP}) defines a projective representation
of the abelian group $\mathbb{T}_d \times \mathbb{T}_d$. It is
therefore clear that
\begin{equation}\label{R-U}
    \mathbb{Z}_d \times \mathbb{Z}_d \ni (m,n) \ \longrightarrow\ U_{mn} \in \mathcal{L}(M_d,M_d) \ ,
\end{equation}
with
\begin{equation}\label{}
    U_{mn} a := u_{mn} a\, u_{mn}^* \ , \ \ \ \  a \in M_d\ ,
\end{equation}
defines the representation of $\mathbb{T}_d \times \mathbb{T}_d$ in
the space of superoperators $\mathcal{L}(M_d,M_d)$.

Now, for any
\begin{equation}\label{}
    a \ : \ \mathbb{Z}_d \ot \mathbb{Z}_d \ \longrightarrow\
    \mathbb{C}\ ,
\end{equation}
let us define a linear map $A \in \mathcal{L}(M_d,M_d)$
\begin{equation}\label{a-A}
 A  = \sum_{m,n=0}^{d-1} a(m,n) U_{n,-m} \ ,
\end{equation}
that is, we define a representation of $M_d$ in
$\mathcal{L}(M_d,M_d)$.

\begin{Proposition}
If $a(m,n) \in \mathbb{R}$, then $A$ is self-adjoint, that is
\begin{equation}\label{}
    A\, x^* = (A\, x)^*\ , \ \ \ \ x \in M_d \ .
\end{equation}
If $a(m,n) \geq 0$, then $A$ is completely positive. If moreover
$\sum_{m,n} a(m,n)=1$, then $A$ is trace preserving and unital.
\end{Proposition}
One proves the following

\begin{Proposition} \label{P-abc}
Let $a,b,c \in M_d$ be represented by $A,B,C \in
\mathcal{L}(M_d,M_d)$, respectively, that is
\[ A  = \sum_{m,n=0}^{d-1} a(m,n) U_{n,-m} \ ,\ \ \ \
B  = \sum_{m,n=0}^{d-1} b(m,n) U_{n,-m} \ ,\ \ \ \ C=
\sum_{m,n=0}^{d-1} c(m,n) U_{n,-m} \ .  \] Then $A \circ B = C$ if
and only if $c = a \ast b$.
\end{Proposition}
Hence, the set of maps constructed via (\ref{a-A}) defines a
commutative subalgebra in $\mathcal{L}(M_d,M_d)$.

\begin{Proposition}
The spectral properties of the linear map (\ref{a-A}) are
characterized by
\begin{eqnarray}\label{}
    A\, u_{kl} &=& \widetilde{a}_{kl}\, u_{kl}\ , \\
    A^\#\, u^*_{kl} &=& \widetilde{\overline{a}}_{kl}\, u^*_{kl}\ ,
\end{eqnarray}
and hence its spectral decomposition reads as follows
\begin{equation}\label{}
    A\ = \sum_{m,n=0}^{d-1} \widetilde{a}(m,n)\, P_{mn} \ ,
\end{equation}
where $P_{mn}$ is a projector defined by
\begin{equation}\label{}
 P_{mn}\, x = \frac 1d\, u_{mn} \, {\rm tr}(u^*_{mn} x) \ ,
\end{equation}
 for any $x \in M_d$.
\end{Proposition}
In particular, if $a(m,n)$ is real, i.e. $A$ is self-adjoint, then
one has
\begin{eqnarray}\label{}
    A\, u_{kl} &=& \widetilde{a}_{kl}\, u_{kl}\ , \\
    A^\#\, u^*_{kl} &=& \widetilde{{a}}_{kl}\, u^*_{kl}\ .
\end{eqnarray}
Note, that the action of $A$ upon the basis $e_{ij}$ is given by
\begin{equation}\label{}
    A\, e_{ij} = \sum_{m,n=0}^{d-1} a(m,n) \lambda^{n(i-j)} \,
    e_{i-m,j-m} \ .
\end{equation}
Hence, diagonal elements satisfy define an invariant subspace in
$M_d$
\begin{equation}\label{}
    A\, e_{ii} = \sum_{m,n=0}^{d-1} a(m,n) \,
    e_{i-m,i-m} \ .
\end{equation}


Let $P_{t,t_0} : \mathbb{Z}_d \times \mathbb{Z}_d \rightarrow [0,1]$
satisfy the following inhomogeneous master equation
\begin{equation}\label{}
    \dot{P}_{t,t_0} = a_t \ast P_{t,t_0}\ , \ \ \ \  \ P_{t_0,t_0}=e\
    .
\end{equation}
Now, following (\ref{a-A}), let us define
\begin{equation}\label{}
    A_{t,t_0} = \sum_{m,n=0}^{d-1} P_{t,t_0}(m,n)\, U_{n,-m} \ ,
\end{equation}
and
\begin{equation}\label{}
    \mathcal{L}_{t} = \sum_{m,n=0}^{d-1} a_{t}(m,n)\, U_{n,-m} \ .
\end{equation}
Then, Proposition \ref{P-abc} implies the following local master
equation for the dynamical map $A_{t,t_0}$:
\begin{equation}\label{Q-M}
    \dot{A}_{t,t_0} = \mathcal{L}_t A_{t,t_0}\ , \ \ \ \  \ A_{t_0,t_0}={\rm id}\
    .
\end{equation}
Note, that the time dependent Markovian generator may be rewrite as
follows
\begin{equation}\label{}
    \mathcal{L}_t\, \rho\, =\, \frac 12 \ {\sum_{m,n}}' a_t(m,n) \Big( [u_{n,-m},\rho
    u_{n,-m}^*] + [u_{n,-m}\rho, u_{n,-m}^*] \Big) \ ,
\end{equation}
where ${\sum}'_{m,n} X_{mn} := \sum_{m,n} X_{mn} - X_{00}$. Hence,
recalling that $a_t(m,n) \geq 0$ for $(m,n) \neq (0,0)$, the above
formula provides the Lindblad form of $\mathcal{L}_t$. The
corresponding spectral representation of the generator reads as
follows
\begin{equation}\label{}
    \mathcal{L}_t = \sum_{m,n=0}^{d-1} \widetilde{a}_t(m,n)\,
    P_{mn}\ .
\end{equation}
Note, that due to $\widetilde{a}_t(0,0)=0$, one has $\mathcal{L}_t
\mathbb{I}_d = 0$.  The corresponding solution of (\ref{Q-M}) is
therefore given by
\begin{equation}\label{}
    A_{t,t_0} = \sum_{m,n=0}^{d-1} \exp\left( \int_{t_0}^t
    \widetilde{a}_u(m,n)\, du \right) \,
    P_{mn}\ .
\end{equation}
If $P_{t,t_0}$ satisfies non-Markovian classical master equation
\begin{equation}\label{}
    \dot{P}_{t,t_0} = a_{t-t_0} \ast P_{t,t_0}\ , \ \ \ \  \ P_{t_0,t_0}=e\
    ,
\end{equation}
then the quantum dynamical map $A_{t,t_0}$ satisfies non-Markovian
equation
\begin{equation}\label{Q-nM}
    \dot{A}_{t,t_0} = \mathcal{L}_{t-t_0} A_{t,t_0}\ , \ \ \ \  \ A_{t_0,t_0}={\rm id}\
    ,
\end{equation}
with the solution given by the following formula
\begin{equation}\label{}
    A_{t,t_0} = \sum_{m,n=0}^{d-1} \exp\left( \int_{0}^{t-t_0}
    \widetilde{a}_u(m,n)\, du \right) \,
    P_{mn}\ .
\end{equation}
This spectral representation of $A_{\tau} := A_{t_0+\tau,t_0}$
enables one to construct the corresponding memory kernel
$\mathcal{K}_\tau$. Using the following representation
\cite{KR-last}
\begin{equation}\label{}
    A_{\tau} = {\rm id} + \int_0^{\tau} F_s ds \ ,
\end{equation}
where
\begin{equation}\label{}
    F_s = \mathcal{L}_s A_s \ ,
\end{equation}
one finds the spectral representation for the super-operator
function $F_s$:
\begin{equation}\label{}
    F_t = \sum_{m,n} f_t(m,n) P_{mn} \
    ,
\end{equation}
with
\begin{equation}\label{f_mn}
    f_t(m,n) = \widetilde{a}_t(m,n) \exp\left( \int_0^t
    \widetilde{a}_u(m,n) du \right) \ .
\end{equation}
Therefore, one may write the corresponding non-local equation
\begin{equation}\label{}
    \dot{A}_t = \int_0^t \mathcal{K}_{t-u} A_u du \ ,
\end{equation}
with the memory kernel is defined in terms of its Laplace transform
as follows
\begin{equation}\label{}
    \widehat{\mathcal{K}}_s = \sum_{m,n} \frac{s
    \widehat{f}_s(m,n)}{1+ \widehat{f}_s(m,n)} \, P_{mn} \ ,
\end{equation}
where $\widehat{f}_s(m,n)$ denotes the Laplace transform of
$f_t(m,n)$. Note, that in general one is not able to invert the
Laplace transform $ \widehat{\mathcal{K}}_s$ and hence the above
formula in general does not have any practical meaning.

\section{Dynamics of composite quantum systems}

Consider now a quantum dynamics of $N$-partite $d$-level quantum
systems defined by
\begin{equation}\label{}
    A_{t,t_0} = \sum_{\mathbf{m},\mathbf{n}\in \mathbb{Z}^N_d}
    P_{t,t_0}(\mathbf{m},\mathbf{n}) \, U_{\mathbf{n},-\mathbf{m}}\
    ,
\end{equation}
where
\begin{equation}\label{}
U_{\mathbf{k},\mathbf{l}} \, x =  u_{\mathbf{k},\mathbf{l}} \, x\,
u_{\mathbf{k},\mathbf{l}} ^*\ ,
\end{equation}
for $x \in M_d^{\ot N}$, and
\begin{equation}\label{}
    u_{\mathbf{k},\mathbf{l}} = u_{k_1,l_1} \ot \ldots \ot u_{k_N,l_N}\
    .
\end{equation}

\begin{Proposition}
Matrices $u_{\mathbf{m},\mathbf{n}}$ satisfy
\begin{eqnarray}
  u_{\mathbf{m},\mathbf{n}} u_{\mathbf{r},\mathbf{s}} &=&
  \lambda^{\mathbf{m}\,\mathbf{s}}\, u_{\mathbf{m}+\mathbf{r},\mathbf{n}+\mathbf{s}} \ , \\
  u^*_{\mathbf{m},\mathbf{n}} &=& \lambda^{\mathbf{m}\,\mathbf{n}}\, u_{-\mathbf{m},-\mathbf{n}} \ ,
\end{eqnarray}
and the following orthogonality relations
\begin{equation}\label{}
    {\rm tr}( u_{\mathbf{m},\mathbf{n}}^* u_{\mathbf{k},\mathbf{l}}) =
    d^N\, \delta_{\mathbf{m},\mathbf{k}} \delta_{\mathbf{n},\mathbf{l}}\ .
\end{equation}

\end{Proposition}
The spectral representation of $A_{t,t_0}$ has the following form
\begin{equation}\label{}
    A_{t,t_0} =  \sum_{\mathbf{m},\mathbf{n}\in \mathbb{Z}^N_d}
    \widetilde{P}_{t,t_0}(\mathbf{m},\mathbf{n})\, P_{\mathbf{m},\mathbf{n}} \ ,
\end{equation}
where $P_{\mathbf{m},\mathbf{n}}$ is a projector defined by
\begin{equation}\label{}
 P_{\mathbf{m},\mathbf{n}}\, x = \frac{1}{d^N}\, u_{\mathbf{m},\mathbf{n}} \, {\rm tr}(u^*_{\mathbf{m},\mathbf{n}} x) \ ,
\end{equation}
 for any $x \in M_d^{\ot N}$. Assuming that
\begin{equation}\label{}
    P_{t,t_0}\ :\ \mathbb{Z}_d^N \times \mathbb{Z}_d^N \
    \longrightarrow\ [0,1]\ ,
\end{equation}
satisfies classical Markovian inhomogeneous master equation
\begin{equation}\label{}
    \dot{P}_{t,t_0} = a_t \ast P_{t,t_0}\ , \ \ \ \  \ P_{t_0,t_0}=e\
    ,
\end{equation}
one obtains
\begin{equation}\label{Q-M}
    \dot{A}_{t,t_0} = \mathcal{L}_t A_{t,t_0}\ , \ \ \ \  \ A_{t_0,t_0}={\rm id}\
    ,
\end{equation}
where the time dependent Markovian generator is defined by
\begin{equation}\label{}
    \mathcal{L}_t =  \sum_{\mathbf{m},\mathbf{n}\in \mathbb{Z}^N_d}
    \widetilde{a}_{t}(\mathbf{m},\mathbf{n})\, P_{\mathbf{m},\mathbf{n}} \
    .
\end{equation}
Hence, the corresponding solution reads as follows
\begin{equation}\label{}
    A_{t,t_0} =  \sum_{\mathbf{m},\mathbf{n}\in \mathbb{Z}^N_d}
    \exp\left( \int_{t_0}^t \widetilde{a}_{u}(\mathbf{m},\mathbf{n})\, du \right) \, P_{\mathbf{m},\mathbf{n}} \
    .
\end{equation}

\section{Commutative dynamics of 2-level system}
\setcounter{equation}{0}

Consider the time dependent generator for a 2-level system defined
by
\begin{eqnarray}  \label{L-2}
  \mathcal{L}_t\rho = - \frac i2 \varepsilon(t) [\sigma_3,\rho] +
  \gamma(t) \Big(\mu \mathcal{L}_1 + (1-\mu) \mathcal{L}_2 \Big) \rho
  + \frac 12 \sum_{\alpha,\beta=0}^1 c_{\alpha\beta}(t) \Big(
   [\pi_\alpha,\rho\pi_\beta]  + [\pi_\alpha\rho,\pi_\beta] \Big) \ ,
\end{eqnarray}
where the time independent Markovian generators $\mathcal{L}_1$ and
$\mathcal{L}_2$ are defined as follows
\begin{eqnarray*}
  \mathcal{L}_1\rho &=& \sigma^+ \rho \sigma^- - \frac 12 \{ \sigma^-\sigma^+,\rho\} \ , \\
  \mathcal{L}_2\rho &=& \sigma^- \rho \sigma^+ - \frac 12 \{ \sigma^+\sigma^-,\rho\} \ . \\
\end{eqnarray*}
One easily shows that
\begin{equation}\label{}
    [ \mathcal{L}_t,\mathcal{L}_s] = 0 \ ,
\end{equation}
and hence $\mathcal{L}_t$ does generate a commutative quantum
dynamics. In (\ref{L-2}) the `mixing' parameter $\mu \in [0,1]$, and
projectors $\pi_\alpha$ are defined by
\begin{equation}\label{}
    \pi_0 = \sigma^-\sigma^+ \ , \ \ \ \ \pi_1 = \sigma^+\sigma^-\ .
\end{equation}
Note, that if $\gamma(t) > 0$ and the time dependent matrix
$||c_{\alpha\beta}(t)||$ is semi-positive definite, than
$\mathcal{L}_t$ defines time dependent Markovian generator. If
\begin{equation}\label{}
    \int_0^t \gamma(u)du > 0 \ ,
\end{equation}
and the matrix
\begin{equation}\label{}
   || \int_0^t c_{\alpha\beta}(u)du \ || \geq 0 \ ,
\end{equation}
for all $t \geq 0$, then $\mathcal{L}_t$ generates non-Markovian
dynamics.

 One easily solves
the corresponding spectral problem for $\mathcal{L}_t$
\begin{eqnarray*}
  \mathcal{L}_t\,\omega \ \, &=& 0 \ , \\
  \mathcal{L}_t\,\sigma^+ &=& \Gamma(t)\, \sigma^+ \ , \\
  \mathcal{L}_t\,\sigma^- &=& \overline{\Gamma(t)}\, \sigma^- \ ,\\
  \mathcal{L}_t\,\sigma_3 \ &=& - \gamma(t)\, \sigma_3 \ ,
\end{eqnarray*}
where the invariant state $\omega$ reads as follows
\begin{equation}\label{}
    \omega = \mu \pi_1 + (1-\mu) \pi_0 \ ,
\end{equation}
and the complex eigenvalue $\Gamma(t)$ is defined by
\begin{equation}\label{}
    \Gamma(t) = - \frac 12 \Big[\, \gamma(t) + c_{00}(t) + c_{11}(t) - 2 c_{10}(t) +
    2i\varepsilon(t)\, \Big] \ .
\end{equation}
Similarly, one solves for the dual generator
\begin{eqnarray*}
  \mathcal{L}_t^\#\mathbb{I}_2\ \, &=& 0 \ , \\
  \mathcal{L}_t^\#\sigma^+ &=& \overline{\Gamma(t)} \sigma^+ \ , \\
  \mathcal{L}_t^\#\sigma^- &=& \Gamma(t) \sigma^- \ ,\\
  \mathcal{L}_t^\#\sigma\ \,&=&  -\gamma(t)\, \sigma \ ,
\end{eqnarray*}
where
\begin{equation}\label{}
    \sigma = (1-\mu)\pi_1 - \mu \pi_0 = \frac 12 \Big( \sigma_3 - \mathbb{I}_2 {\rm tr}(\omega
    \sigma_3) \Big) \ .
\end{equation}
Hence, introducing a bi-orthogonal basis
\begin{eqnarray*}
  g_0 &=& \omega \ , \ \ \  \ \ \ \, h_0 \,=\, \mathbb{I}_2 \ , \\
  g_1 &=& \sigma^+  ,\ \ \ \ \ \ h_1 \,=\, \sigma^+ , \\
  g_2 &=& \sigma^-  , \ \ \ \ \  \ h_2 \,=\, \sigma^- , \\
  g_3 &=& \sigma_3 \ , \ \ \ \ \ \ h_3 \,=\, \sigma \ ,
\end{eqnarray*}
such that
\begin{equation}\label{}
    (g_\alpha,h_\beta) = {\rm tr}(g_\alpha^* h_\beta) =
    \delta_{\alpha\beta}\ ,
\end{equation}
one has
\begin{equation}\label{}
    \mathcal{L}_t \rho = \sum_{\alpha=0}^3 \lambda_\alpha(t)\,
    g_\alpha\, {\rm tr}(h_\alpha^*\, \rho) \ ,
\end{equation}
with
\begin{equation}\label{}
    \lambda_0(t)=0\ , \ \ \ \ \lambda_1(t) = \overline{\lambda_2(t)} = \Gamma(t)
    \ , \ \ \ \ \lambda_3(t) = - \gamma(t)\ .
\end{equation}
Hence, the solution to the Markovian master equation
\begin{equation}\label{}
    \dot{A}_{t,t_0} = {\cal L}_t A_{t,t_0}\ , \ \ \ A_{t_0,t_0} =
    {\rm id} \ ,
\end{equation}
reads
\begin{equation}\label{}
    A_{t,t_0}\, \rho = \sum_{\alpha=0}^3 \exp\left( \int_{t_0}^t
    \lambda_\alpha(u)\, du \right) \,
    g_\alpha\, {\rm tr}(h_\alpha^*\, \rho) \ .
\end{equation}
Consider now
\begin{equation}\label{}
    V \ :\ M_2 \ \longrightarrow\ M_2 \ ,
\end{equation}
defined by
\begin{eqnarray}\label{}
    V\, a &=&  e_{00} \Big( \mu\, {\rm tr}( e_{11} a) +  {\rm tr}( e_{00}
          a) \Big) + e_{11} \Big(  (1-\mu)\, {\rm tr}(
    e_{11}a) - {\rm tr}( e_{00} a) \Big) \nonumber \\
          &+&  e_{10} {\rm tr}( e_{01} a) + e_{01} {\rm tr}( e_{10}
          a)\ .
\end{eqnarray}
One easily finds for the inverse
\begin{eqnarray}\label{}
    V^{-1} a &=&  e_{00} \Big( - \mu\, {\rm tr}( e_{11} a) + (1-\mu)\, {\rm tr}( e_{00}
          a) \Big) + e_{11} \Big(  {\rm tr}(
    e_{11}a) + {\rm tr}( e_{00} a) \Big) \nonumber \\
          &+&  e_{10} {\rm tr}( e_{01} a) + e_{01} {\rm tr}( e_{10}
          a)\ ,
\end{eqnarray}
and hence
\begin{eqnarray}\label{}
    V^{-1\#} a &=&  e_{00} \Big(  {\rm tr}( e_{11} a) + (1-\mu)\, {\rm tr}( e_{00}
          a) \Big) + e_{11} \Big(  {\rm tr}(
    e_{11}a) -\mu\, {\rm tr}( e_{00} a) \Big) \nonumber \\
          &+&  e_{10} {\rm tr}( e_{01} a) + e_{01} {\rm tr}( e_{10}
          a)\ .
\end{eqnarray}
 One finds
\begin{eqnarray}
  V\, e_{00} =  \sigma_3\ , \ \ \   V\, e_{11} =  \omega \ , \ \ \   V\, \sigma^\pm = \sigma^\pm\ ,
\end{eqnarray}
and
\begin{eqnarray}
  V^{-1\#} e_{00} = \sigma\ , \ \ \   V^{-1\#} e_{11} = \mathbb{I}_2 \ ,
  \ \ \   V^{-1\#} \sigma^\pm = \sigma^\pm\ .
\end{eqnarray}
Hence, defining
\begin{equation}\label{}
    f_0 = e_{11}\ , \ \ \ f_1 = \sigma^+\ , \ \ \ f_2 = \sigma^-\ ,
    \ \ \ f_3 = e_{00} \ ,
\end{equation}
one has
\begin{equation}\label{}
    g_\alpha = V\, f_\alpha\ ,
    \ \ \ \ \ h_\alpha = V^{-1\#} f_\alpha\ ,
\end{equation}
which shows that $V$ diagonalizes $\mathcal{L}_t$ and $A_{t,t_0}$,
that is,
\begin{equation}\label{}
    \mathcal{L}_t = \sum_{\alpha=0}^3 \lambda_\alpha(t) V P_\alpha
    V^{-1} \ ,
\end{equation}
and
\begin{equation}\label{}
    A_{t,t_0} = \sum_{\alpha=0}^3 \exp\left( \int_{t_0}^t
    \lambda_\alpha(u)\, du \right) \, V P_\alpha
    V^{-1}      \ ,
\end{equation}
where
\begin{equation}\label{}
    P_\alpha \rho = f_\alpha {\rm tr}(f_\alpha^* \rho)\ .
\end{equation}

\section{Conclusions}

In this paper we analyzed a class of commutative dynamics of quantum
open systems. It is shown that such evolution may be effectively
described via spectral analysis of the corresponding time dependent
generators. The characteristic feature of the corresponding
time-dependent dynamical map is that all its eigenvectors do not
depend on time (only its eigenvalues do). Actually, majority of
examples studied in the literature (see e.g. \cite{OPEN}) belong to
this class. If eigenvectors vary in time then the solution is
formally defined by the time ordered exponential but the problem of
finding an explicit solution is rather untractable. We stress that
both Markovian and non-Markovian dynamics were studied. Our analysis
shows that the local approach to non-Markovian dynamics proposed in
\cite{PRL} is much more suitable in practice than the corresponding
non-local approach based on the memory kernel.

\section*{Acknowledgments}
This work was partially supported by the Polish Ministry of Science
and Higher Education Grant No 3004/B/H03/2007/33.


\begin{thebibliography}{1} \bibliographystyle{plain}

\bibitem{OPEN} H.-P. Breuer and F. Petruccione, {\em The Theory of Open
Quantum Systems}, (Oxford Univ. Press, Oxford, 2007).

\bibitem{QIT} M. A. Nielsen and I. L. Chuang, {\it Quantum
Computation and Quantum Information} (Cambridge Univ. Press,
Cambridge, 2000).

\bibitem{Gorini} V. Gorini, A. Kossakowski, and E.C.G. Sudarshan, J. Math. Phys.
{\bf 17}, 821 (1976).

\bibitem{Lindblad} G. Lindblad, Comm. Math. Phys. {\bf 48}, 119 (1976).

\bibitem{Alicki} R. Alicki and K. Lendi, {\em Quantum Dynamical Semigroups
and Applications},  (Springer, Berlin, 1987).




\bibitem{Wilkie} J. Wilkie,  Phys. Rev. E {\bf 62},  8808 (2000); J. Wielkie and Yin Mei
Wong, J. Phys. A: Math. Theor. {\bf 42}, 015006 (2009).

\bibitem{Budini} A. A. Budini, Phys. Rev. A {\bf 69}, 042107 (2004);
\textit{ibid.} {\bf 74}, 053815 (2006).

\bibitem{B-2004} H.-P. Breuer,  Phys. Rev. A {\bf 69} 022115 (2004);
\textit{ibid.} {\bf 70}, 012106 (2004).

\bibitem{Wodkiewicz}  S. Daffer, K. W\'odkiewicz, J.D. Cresser, and J.K.
McIver, Phys. Rev. A {\bf 70}, 010304 (2004).

\bibitem{Lidar} A. Shabani and D.A. Lidar, Phys. Rev. A {\bf 71}, 020101(R)
(2005).

\bibitem{Maniscalco1} S. Maniscalco,  Phys. Rev. A {\bf 72},  024103
(2005).

\bibitem{Maniscalco2} S. Maniscalco and F.  Petruccione,  Phys. Rev. A {\bf 73},
012111 (2006).

\bibitem{Maniscalco-09} J. Piilo, K. Harkonen, S. Maniscalco, K.-A.
Suominen, Phys. Rev. Lett. {\bf 100}, 180402 (2008); Phys. Rev. A
{\bf 79}, 062112 (2009).


\bibitem{KR} A. Kossakowski and R. Rebolledo,  Open Syst. Inf. Dyn.
{\bf 14}, 265 (2007); \textit{ibid.} {\bf 15}, 135 (2008).

\bibitem{KR-last} A. Kossakowski and R. Rebolledo, Open Syst. Inf.
Dyn. {\bf 16}, 259 (2009).

\bibitem{B} H.-P. Breuer and B. Vacchini, Phys. Rev. Lett. {\bf 101} (2008) 140402;
Phys. Rev. E {\bf 79}, 041147 (2009).

\bibitem{PRA-Saverio} D. Chru\'sci\'nski, A. Kossakowski and S. Pascazio, Phys. Rev.
A {\bf 81}, 032101 (2010).


\bibitem{PRL} D. Chru\'sci\'nski andA. Kossakowski, Phys. Rev. Lett.
{\bf 104}, 070406 (2010).




\bibitem{damping} H.-J. Briegel and B.-G. Englert, Phys. Rev. A {\bf 47},
3311 (1993).



\bibitem{Gantmacher} F.R. Gantmacher, {\em The Theory of Matrices}, Chelsea Publishing Co.,
NY 1960.

\end{thebibliography}
\end{document}